\documentstyle[pra,graphicx,aps]{revtex}
  
\begin{document} 
\title{ Non-linear properties of strongly pumped lasers }

\author{ Gennady Koganov\thanks{quant@bgumail.bgu.ac.il} and 
Reuben Shuker}
\address{ Department of Physics, Ben Gurion University of the Negev, 84105
Beer Sheva, Israel}
\maketitle
\begin{abstract}
Bloch equations for the atomic population and the polarization/coherence 
and the equation of motion for the photon number in a laser are solved in 
steady state as a function of the pump rate. Two level atom and two modes 
of three levels atom are investigated. Close to 
threshold the usual linear dependence of the intensity on the pump rate in 
found for all cases. However, far above threshold strongly nonlinear 
dependence is encountered. In the cases for which the pump connects the 
lower lasing state to one of the excited states the character of the 
non-linearity differs crucially from the cases when the pump in not 
related directly to the lower lasing state. Non-monotonic dependence of 
laser intensity upon the pump rate is predicted. Detailed discussion of 
the nonlinear behavior is presented, including saturation and depletion 
effects.
\end{abstract}

The behavior of the number of photons and the laser intensity as function 
of the pump rate near and above threshold has recently been studied 
[\ref{Carmicael}]. In particular, linear dependence of the intensity close 
to threshold and some non-linearity above threshold has been found [\ref{Smith,Gardiner}, 
\ref{Hart, Kennedy}]. The purpose of this paper is to point out the 
non-linearity of the laser intensity as function of pump and the crucial 
role of the rate of depletion of the lower lasing state in the dynamics of 
lasers. It is usually believed [\ref{Sigman}] that the faster is the rate of 
depletion of the lower laser level, the better, because this gives 
rise to a larger inversion at the lasing transition. However, as it is 
shown in this paper, there is a limitation on the depletion rate of 
the lower lasing level. For any arbitrarily large pump rate there 
exists a critical value of the depletion rate of the lower lasing 
level, at which the lasing is broken down.\\

In this paper we study closed two and three level systems with incoherent pump.
Consider first two level scheme shown on Fig. 1. The corresponding optical 
Bloch equations are

\begin{eqnarray}
\dot{\rho}_{11}&=& -\gamma \rho_{11} + \Gamma \rho_{00} -ig(z\rho_{01} - 
c.c.),
\label{eq1}\\
\dot{\rho}_{10}&=& -\gamma_{\perp} \rho_{10} + igz(\rho_{11} - \rho_{00}),
\label{eq2}\\
\dot{n}&=& -2\kappa n + iNg(z\rho_{01} - c.c.),
\label{eq3}
\end{eqnarray}

where $N$ is the total number of atoms, $g$ is the electric dipole 
coupling, $\kappa$ is cavity decay rate, $n=z^{*}z$ is the number of 
photons inside the cavity, $\gamma$ and $\Gamma$ are spontaneous emission 
rate and the pump rate respectively. The steady state solution for the 
photon number $n$ reads

\begin{equation}
n=\frac{N\gamma}{4\kappa}[P-1-(P+1)\frac{\kappa \gamma_{\perp}}{Ng^{2}}],
\label{eq4}
\end{equation}

where $P\equiv \Gamma/\gamma$ is the relative pump rate. It follows from  
eq. (\ref{eq4}) that in order to get lasing the following condition must 
be 
satisfied

\begin{equation}
P>P_{thr}=\frac{1+\frac{\kappa\gamma_{\perp}}{Ng^{2}}}{1-
\frac{\kappa\gamma_{\perp}}{Ng^{2}}},
\label{eq5}
\end{equation}

which is usually referred to as a threshold condition. The necessary 
condition for eq. (\ref{eq5}) to be true is

\begin{equation}
\gamma_{\perp}<\frac{Ng^{2}}{\kappa}.
\label{eq6}
\end{equation}

If this inequality is not satisfied,  lasing cannot be obtained at any 
pump 
rate. In case of $Ng^{2}/\kappa\gamma_{\perp} \gg 1$, when eq. (\ref{eq6}) 
is satisfied automatically, the threshold condition is simple

\begin{equation}
P >1
\label{eq7}
\end{equation}

which means that the pump rate $\Gamma$ should exceed the decay rate 
$\gamma$ of the upper lasing level. Inequality (\ref{eq6}) gives rise to a restriction on the pump rate. Indeed, it follows from the
derivation of the relaxation rates (see Refs. [\ref{derivation of 
gammas}]) that

\begin{equation}
\gamma_{\perp} = \frac{\Gamma + \gamma + \gamma_{ph}}{2}
\label{eq8}
\end{equation}

where $\gamma_{ph}$ stands for the contribution from the phase destruction 
processes like elastic collisions. Combining eqs. (\ref{eq6})-
(\ref{eq8}) we obtain the following restriction for the pump rate $P$

\begin{equation}
1< P <\frac{2Ng^{2}}{\kappa \gamma} -1 - \delta, \;\;\;\;\;\;\;\; 
\delta \equiv \frac{\gamma_{ph}}{\gamma}.
\label{eq9}
\end{equation}

At first glance at eq. (\ref{eq4}), the dependence of the mean photon 
number $n$ upon the pump parameter $P$ seems to be linear. However this is 
not true due to the dependence of the transversal relaxation rate 
$\gamma_{\perp}$ upon the pump parameter. Indeed, taking into account 
eq. (\ref{eq8}) we get, instead of eq. (\ref{eq4}), the following 
expression for the mean photon number 

\begin{equation}
n = \frac{N\gamma}{4\kappa }[P -1 - (P+1)(P+1+\delta) 
\frac{\kappa \gamma}{2Ng^{2}}].
\label{eq10}
\end{equation}

The dependence (\ref{eq10}) is not linear, it is rather quadratic in $P$ 
(see Fig. 2 ) with maximum at 

\begin{equation}
P_{0}=\frac{Ng^{2}}{\kappa\gamma} -1 - \frac{\delta}{2}.
\label{eq11}
\end{equation}

Note that condition (\ref{eq9}) is necessary but not enough to get lasing. 
The region of the pump parameter $P$ in which lasing is really  possible is given 
by (see fig. 2)

\begin{equation}
1 < P < \frac{2Ng^{2}}{\kappa \gamma} - \delta - 3,
\label{eq12}
\end{equation}

Here we assumed that $Ng^{2}/\kappa\gamma \gg 1$ which is normally true. 
Maximum value of the photon number is defined by eqs. (\ref{eq10}), 
(\ref{eq11}) and has the following form, provided $Ng^{2}/\kappa\gamma \gg 
1$ 

\begin{equation}
n_{max} \approx \frac{1}{8}(\frac{Ng}{\kappa})^{2}
\label{eq15}
\end{equation}

The non-linearity in eq. (\ref{eq10}) stems from the fact that according 
to eq. (\ref{eq6}) the rate of transversal relaxation $\gamma_{\perp}$ 
depends upon the pump rate $\Gamma$. The larger is the pump rate, the 
shorter is atomic radiative life time which is inverse proportional to 
$\gamma{perp}$. Thus, a "too fast" pump can give rise 
to a reduction of the  photon number (or intensity) and can even switch 
off the lasing process. At relatively low pump this reduction is 
compensated by the gain which is given by the linear term in eq. 
(\ref{eq10}). However, when the pump parameter  $P$ becomes large enough, 
i.e. of order of $P \sim Ng^{2}/\kappa\gamma$, the non-linearity plays an 
important role and cannot be 
neglected. The two level model exhibits an unusual non-linearity with 
respect to pump rate.\\

To get a more realistic model one has to involve multilevel schemes. In 
such schemes however, the depletion of the lower lasing level is not 
necessarily associated with pumping mechanism and therefore, the character 
of non-linearity may be quite different. \\

Let us consider now a more realistic three level model shown in fig. 3, 
which is a particular case of a general three level model considered in 
Ref. [\ref{our PRA 1990}]. Note that the two schemes depicted in figs. 3a 
and 3b are equivalent mathematically, i.e., they are formally described by 
the same set of equations. However, physically the dynamic properties of 
these two schemes are completely different because of the different ways 
of pumping as described by the role of the relevant parameters (see figs. 
3a and 3b). In the scheme of fig. 3a the lower lasing state $|0>$  is 
depleted  with the rate $\gamma_{02}$ to the ground state $|2>$, and then 
the pumping is used to excite the atom to the upper lasing state $|1>$ 
with pump rate $\gamma_{21}$. In the scheme of fig. 3b the pump mechanism 
is used to excite the atom from the lower lasing state $|0>$ which is now 
the ground state, to the upper lasing state $|2>$ which is then depleted with 
rate $\gamma_{21}$. In the last case the role of the pump rate is played by 
the rate $\gamma_{02}$. As will be seen in the following this difference in 
the pump mechanism is crucial for the dynamics of laser. \\

Let us write down a set of optical Bloch equations for the three level 
model (fig. 3). 

\begin{eqnarray}
\dot{\rho}_{11}&=&\gamma_{21}\rho_{22} - \gamma_{10} \rho_{11} - 
g(iz\rho_{01} 
+ c.c.),
\label{eq2.1}\\
\dot{\rho}_{22}&=&\gamma_{02}\rho_{00} - \gamma_{21} \rho_{22},
\label{eq2.2}\\
\dot{\rho}_{10}&=&- \gamma_{\perp}\rho_{10} + igz (\rho_{11} - \rho_{00}),
\label{eq2.3}\\
\dot{n}&=&-2\kappa n + Ng(iz\rho_{01} + c.c.)
\label{eq2.4}
\end{eqnarray}

The steady state solution for the number of photons inside the cavity reads

\begin{equation}
n=\frac{N}{2\kappa} \frac{\gamma_{21} (\gamma_{02} - 
\gamma_{10})}{\gamma_{02}+ 2\gamma_{21}} - \frac{\gamma_{\perp}}{2g^{2}} 
\frac{\gamma_{02}\gamma_{21} + \gamma_{02}\gamma_{10} + 
\gamma_{21}\gamma_{10}}{\gamma_{02}+ 2\gamma_{21}}.
\label{eq2.5}
\end{equation}

This formulae is valid for both models shown in figs. 3a and 3b. It seems 
from eq. (\ref{eq2.5}) that the photon number $n$ depends upon both rates 
$\gamma_{02}$ and $\gamma_{21}$ in a similar way, i.e. both dependencies 
are of linear-fractional type, which means that $n$ tends asymptotically 
to some fix value when increasing either  the rate $\gamma_{02}$ or 
$\gamma_{21}$. This type of non-linearity has been brought about by the 
nonlinear dependence of both longitudinal relaxation rate and equilibrium 
inversion $\Delta$ upon the rates $\gamma_{02}$ and $\gamma_{21}$. This 
can be shown explicitly by rewriting eq. (\ref{eq2.5}) in the following 
form

\begin{equation}
n = \frac{N\gamma_{\parallel} \Delta}{4\kappa} - 
\frac{\gamma_{\perp}\gamma_{\parallel}}{4g^{2}},
\label{eq2.6}
\end{equation}

where

\begin{eqnarray}
\gamma_{\parallel} \equiv \frac{2(\gamma_{21}\gamma_{02} + 
\gamma_{02}\gamma_{10}+ \gamma_{21}\gamma_{21})}{\gamma_{02}+ 2 
\gamma_{21}}, 
\label{eq2.7a}\\
\Delta \equiv \frac{\gamma_{21}(\gamma_{02}- 
\gamma_{10})}{\gamma_{21}\gamma_{02} + \gamma_{02}\gamma_{10}+ 
\gamma_{21}\gamma_{21}}.
\label{eq2.7}
\end{eqnarray}

However, as it has already been mentioned in the discussion of the two 
level model, the dipole relaxation rate $\gamma_{\perp}$ is related to 
the  other rates. In the three level case under consideration the 
following relation for $\gamma_{\perp}$ takes place

\begin{equation}
\gamma_{\perp}=\frac{1}{2}(\gamma_{10} + \gamma_{02} + \gamma_{ph}),
\label{eq2.8}
\end{equation}

where $\gamma_{ph}$ stands for collisional dephasing rate. Note that 
$\gamma_{\perp}$ depends on $\gamma_{02}$ but does not depend upon the 
rate 
$\gamma_{21}$. It is this asymmetry that is responsible for the difference 
in dynamic properties of lasers with the above two different pump 
mechanisms. Indeed, in the case of the scheme of fig. 3a, $\gamma_{\perp}$ 
does not depend upon the pump rate $\gamma_{21 }$ and hence the dependence 
of the photon number upon pump remains linear-fractional. In contrast, in 
the case of fig. 3b the role of pump rate plays $\gamma_{02}$ and thus 
$\gamma_{\perp}$ brings about additional dependence upon the pump rate. 
Therefore the type of non-linearity of the photon number as function of 
pump in this case differs from that of the scheme 3a. This difference is 
crucial for the dynamics of the lasers with the different pump schemes. \\

Let us consider the two above mentioned modes of pumping separately. In 
the case of the scheme on fig. 3a equations (\ref{eq2.5}) and 
(\ref{eq2.8}) result in the following expression for the photon number

\begin{equation}
n_{1}=\lambda_{1}\frac{P_{1}(1-\epsilon_{1}) - s_{1}(1+\epsilon_{1} + 
\delta_{1})[P_{1}(1+\epsilon_{1}) + \epsilon_{1}]}{1 + 2 P_{1}}
\label{eq2.9}
\end{equation}

where

\begin{equation}
\lambda_{1}\equiv \frac{N\gamma_{02}}{2\kappa}, \;\;\;\; s_{1}\equiv 
\frac{\kappa 
\gamma_{02}}{2N g^{2}},\;\;\;\; \epsilon_{1}\equiv 
\frac{\gamma_{10}}{\gamma_{02}},\;\;\;\; P_{1}\equiv 
\frac{\gamma_{21}}{\gamma_{02}}, \;\;\;\; \delta_{1}\equiv 
\frac{\gamma_{ph}}{\gamma_{02}}.
\nonumber
\end{equation}

In fig. 4a $n_{1}$ is plotted as function of pump parameter $P_{1}$. At 
the beginning, when $P_{1}<<1$, $n_{1}$ grows linearly with the 
pump rate $P_{1}$. However, when the pump is strong enough, i.e. 
$\gamma_{21}\ge \frac{\gamma_{02}}{2}$, the dependence $n_{1}(P_{1})$ 
becomes non-linear. The origin of this non-linearity lies in the large 
disparity between the rates of depletion of the states $|2>$ and $|0>$. 
When the lower lasing state $|2>$ is depleted much faster than the ground 
state $|0>$, there is a bottleneck at the transition $|0>-|2>$ so that 
further increase of the pump rate does not result in increase of the 
photon number. This kind of non-linearity was mentioned in Ref.[\ref{Hart, 
Kennedy}]. \\

Threshold condition for the scheme 3a is

\begin{equation}
P_{1}> \frac{\epsilon_{1}s_{1}(1 + \epsilon_{1} + \delta_{1} )}{1 - 
\epsilon_{1}- 
s_{1}(1+\epsilon_{1} + \delta_{1})(1+ \epsilon_{1})}
\label{eq2.10}
\end{equation}

It follows from eq.(\ref{eq2.10}) that there are two additional necessary 
conditions for getting lasing:

\begin{eqnarray}
\epsilon_{1}&<&1,
\label{eq2.11}\\
s_{1}&<&\frac{1- \epsilon_{1}}{(1+ \epsilon_{1}+ \delta_{1})(1+ 
\epsilon_{1})}.
\label{eq2.12}
\end{eqnarray}

As it follows from the definition of the equilibrium inversion, $\Delta$, 
(see eq.(\ref{eq2.7}) ), the condition (\ref{eq2.11}) merely means that 
there should be initial positive inversion at the lasing transition. As 
for eq.(\ref{eq2.12}) we can rewrite it assumming that 
$\frac{\kappa\gamma_{10}}{2Ng^{2}} << 1$ in the following form

\begin{equation}
1 < \frac{\gamma_{02}}{\gamma_{10}} < \frac{2Ng^{2}}{\kappa \gamma_{10}} - 
\frac{\gamma_{ph}}{\gamma_{10}} - 3
\label{eq2.13}
\end{equation}

which is similar to eq.(\ref{eq12}). We came to the same restriction for 
the rate of depletion of the lower lasing level as for the two level 
model. This time however, this does not result in any restriction for the 
pump rate $\gamma_{21}$, which can be readily understood because in 
contrast to the two level scheme, in the scheme under consideration, there 
is no connection between the transversal relaxation rate and the pump 
rate. Inequality (\ref{eq2.12}) has also another important meaning which 
becomes 
obvious if we rewrite it in the following way

\begin{equation}
N > N_{min} = \frac{\kappa \gamma_{02}}{2g^{2}} \frac{(1+\epsilon_{1} + 
\delta_{1})(1+\epsilon_{1})}{1- \epsilon_{1}}.
\label{eq2.14}
\end{equation}

Thus there is a minimal number of atoms $N_{min}$ needed to get lasing. If 
$N \le N_{min}$ one cannot get lasing at any pump rate. \\

Let us consider the scheme shown in fig. 3b. Now instead of 
eq.(\ref{eq2.9}) we have for 
the photon number

\begin{equation}
n_{2} =\lambda_{2} \frac{P_{2}- \epsilon_{2} - s_{2}(P_{2}+ \epsilon_{2} 
+\delta_{2})(P_{2}+ \epsilon_{2} + P_{2}\epsilon_{2})}{P_{2} + 2},
\label{eq2.15}
\end{equation}

where

\begin{equation}
\lambda_{2}\equiv \frac{N\gamma_{21}}{2\kappa}, \;\;\;\; s_{2}\equiv 
\frac{\kappa 
\gamma_{21}}{2N g^{2}},\;\;\;\; \epsilon_{2}\equiv 
\frac{\gamma_{10}}{\gamma_{21}},\;\;\;\; P_{2}\equiv 
\frac{\gamma_{02}}{\gamma_{21}}, \;\;\;\; \delta_{2}\equiv 
\frac{\gamma_{ph}}{\gamma_{21}}.
\nonumber
\end{equation}

Note that the third term in the numerator contains additional dependence 
upon the relative pump rate $P_{2}$. This additional dependence has been 
brought about by the fact that the dipole relaxation rate $\gamma_{\perp}$ 
depends on the pump rate $\gamma_{02}$. The photon number $n_{2}$ is 
plotted in fig. 4b as a function of the pump rate. 
Again, as in the previous case, the photon number grows linearly at 
relatively weak pump and becomes non-linear at strong pump. However, the 
character of this non-linearity is crucially different. It has the same 
form as in the case of the two level scheme (cf. fig. 2). This difference 
is due to the above mentioned additional dependence of the photon number 
upon the pump.  The photon number has a maximum at 

\begin{equation}
P_{2}=\frac{Ng^{2}}{\kappa \gamma_{21}} - \frac{\delta_{2}}{2} - 
\epsilon_{2}.
\label{eq2.16}
\end{equation}

Here we assummed that $\epsilon_{2}<<1$. Eq.(\ref{eq2.16}) is 
similar to eq.(\ref{eq11}) for the two level model. In order to get lasing 
the following double inequality must be satisfied, provided $s_{2},\epsilon_{2}<<1$

\begin{equation}
1 < P_{2} < \frac{1}{s_{2}}-\delta_{2}
\label{eq2.17}
\end{equation}

which is similar to eq.(\ref{eq12}) for the two level model. Again we got 
the same restriction for the rate of depletion of the lower lasing state 
$|0>$ as in the case of eq.(\ref{eq2.13}). This time however, this is the 
restriction for the pump rate $P_{2}$ since when $P_{2}$ reaches the value 
of $1/s_{2} - \delta_{2}$, the lasing is broken down for the same reason 
as in the case of two level model.\\

In summary, we have shown that the nonlinear dependence of the photon 
number upon the pump parameter is very sensitive to the type of pump 
mechanism. When the pumping is used to excite the lower lasing state to 
one of the upper atomic states, the laser exhibits a peculiar behavior 
shown in figs. 2 and 4b for two and three level schemes respectively. 
The number of photons as a function of pump reaches its maximum and then 
slows down untill some 
critical point at which the lasing ceases. It should be noted that the 
described non-linearity manifests itself at very large pump rates when the 
ratio $p/p_{thr}$ reaches a few orders of magnitude. Such a regime may be difficult to achieve in conventional lasers. For instance, in order 
to reach the non-linear region in experiments with GaAs laser used in Ref. 
[\ref{Jin}] the ratio $p/p_{thr}$ should be of order  of $10^{5}$. Nevertheless, it may be realized in a microlaser. \\


This work was supported in part by the grant from the Israeli Ministry of 
Immigrant 
Absorbtion.


{\bf References}

\begin{enumerate}
\item\label{Carmicael} See for example P.R. Rice and H.J. Carmichael, 
Phys. Rev. A 
{\bf 50}, 4318 (1994) and references therein.
\item\label{Smith,Gardiner} A.M. Smith and C.W. Gardiner, Phys. Rev. A 
{\bf 41}, 
2730 (1990).
\item\label{Hart, Kennedy} D.L. Hart and T.A.B. Kennedy, Phys. Rev. A {\bf 
44}, 4572 
(1991).
\item\label{Sigman} See, for example, A.E. Sigman, Lasers (University 
Science Books, Mill Valley, 1986).
\item\label{derivation of gammas} L.A. Lugiato, Physica {\bf 81A}, 565 
(1975); L.A. 
Pokrovsky, Teoertich. i Matematich. Fizika {\bf 37}, 102 (1978); T. 
Arimitsu and F. 
Shibata, J. Phys. Soc. Jpn. {\bf 52}, 772 (1983).
\item\label{our PRA 1990} A.M. Khazanov, G.A. Koganov and E.P Gordov, 
Phys. Rev. 
A {\bf 42}, 3065  (1990). 
\item\label{Jin} R. Jin, D. Boggavarapu, M. Sargent III, P. Meystre, H.M. 
Gibbs, and G. 
Khitrova, Phys. Rev. A {\bf 49}, 4038(1994).

\end{enumerate}      
\newpage
\includegraphics{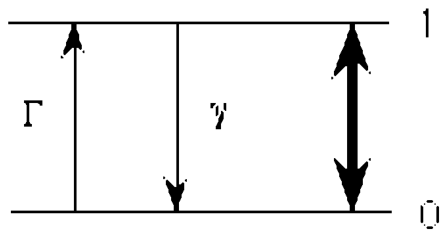}
\includegraphics{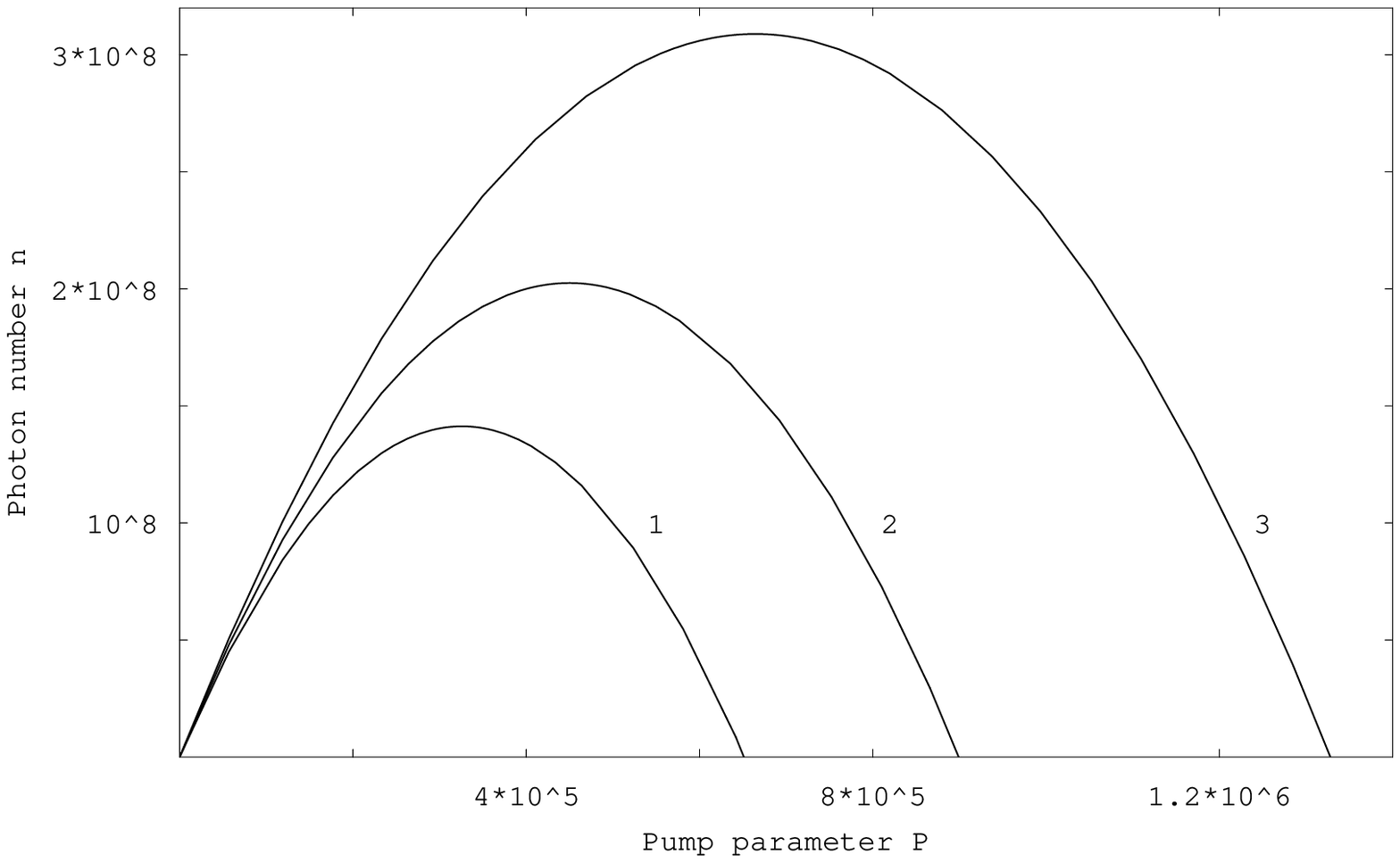}
\includegraphics{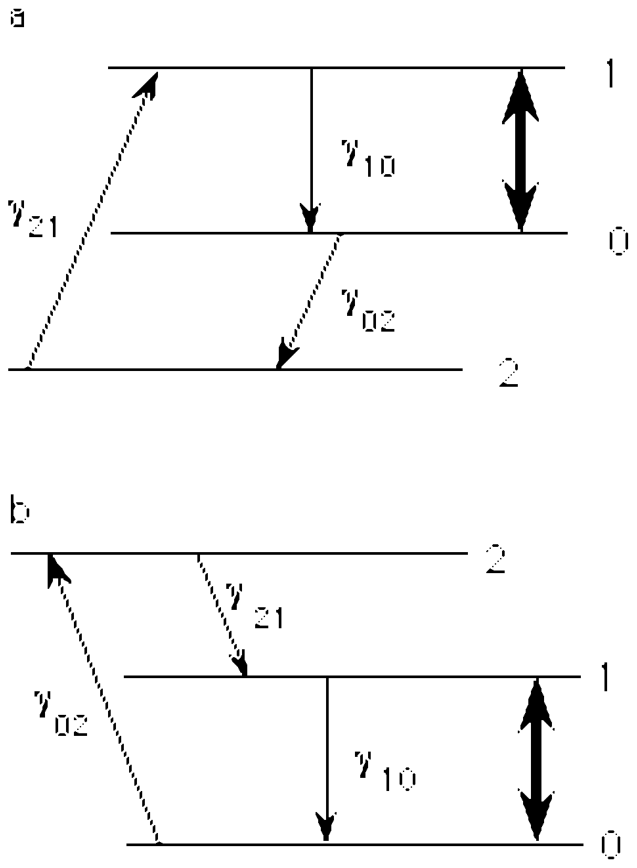}
\includegraphics{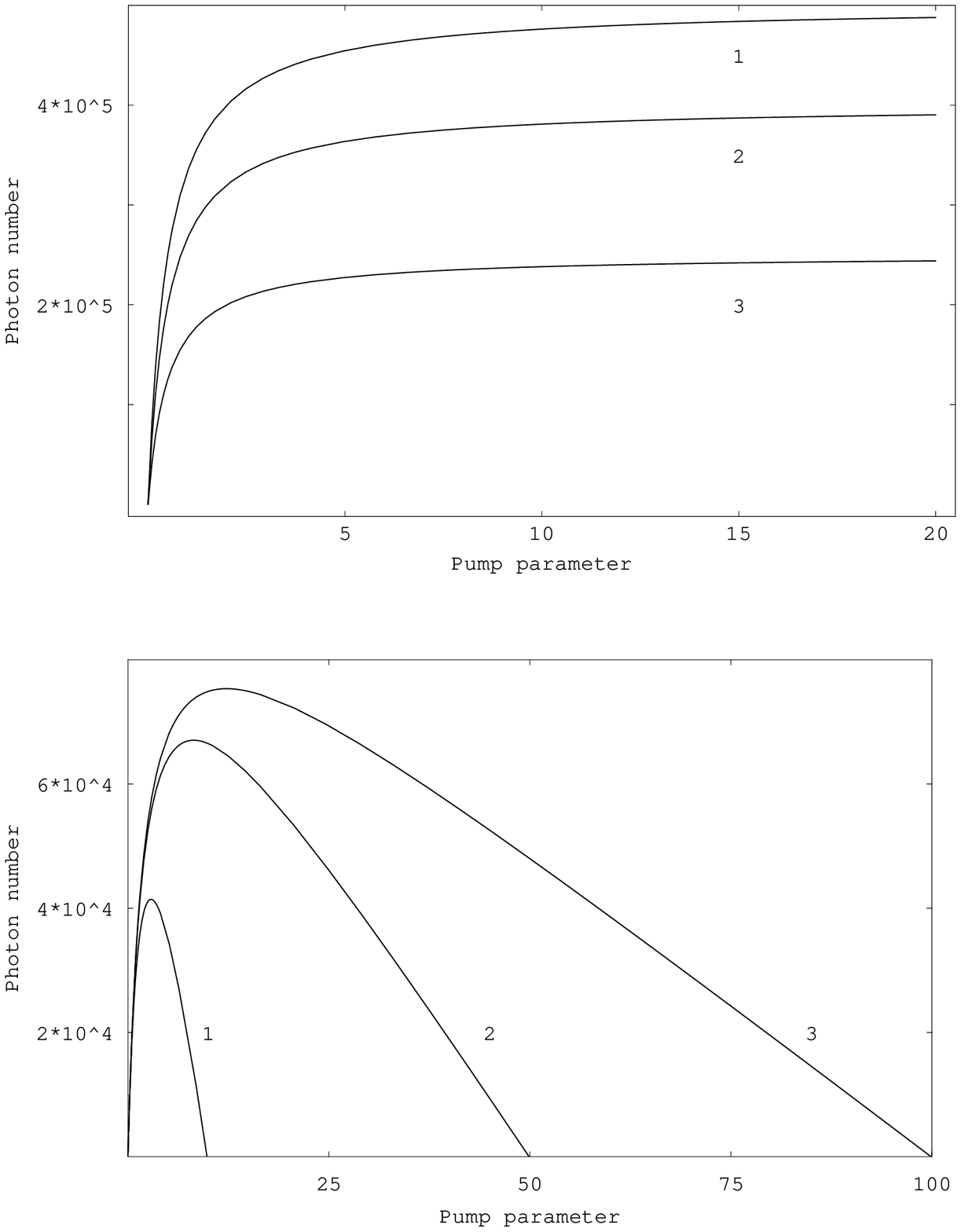}

\newpage
\begin{center}
{\bf Figure captions}
\end{center}
1.  Two level scheme.\\
2. The photon number vs relative pump rate for the two level scheme. The 
parameters are $\lambda = 10^{3}$, $\delta = 10^{5}$, s =  $7\; 10^{-7}$, 
 $10^{-6}$ and  $1.33\; 10^{-6}$ for curves 1, 2 and 3, respectively.\\
3. Three level schemes with different pump mechanism. \\
4. Photon number vs pump rate for the pump schemes shown in figs. 3a and 
3b. The parameters are: $\lambda_{1}=10^{6}$, $\epsilon_{1}=0.01$, 
$\delta_{1}=0$, $s_{1}=$ 0 (curve 1), 0.2 (curve 2) and 0.5 (curve 3) for 
fig 4a; $\lambda_{2}=10^{5}$, $\epsilon_{2}=0$, $\delta_{2}=0.1$, $s_{2}=$ 
0.1 (curve 1), 0.02 (curve 2) and 0.01 (curve 3) for fig 4b.

\end{document}